\documentclass[final, nopreprintline,12pt,3p,sort&compress]{elsarticle}

\usepackage{amssymb}
\usepackage[mathscr]{euscript}
\usepackage{amsthm}
\usepackage{multirow}

\usepackage{amsmath}
\usepackage{array}

\usepackage[hyphens]{url}
\usepackage{caption}
\usepackage{pdfpages}

\usepackage[colorlinks,linkcolor=green, citecolor=green]{hyperref}

\journal{Arxiv}

\begin{document}

\begin{frontmatter}

%% Title, authors and addresses

%% use the tnoteref command within \title for footnotes;
%% use the tnotetext command for theassociated footnote;
%% use the fnref command within \author or \address for footnotes;
%% use the fntext command for theassociated footnote;
%% use the corref command within \author for corresponding author footnotes;
%% use the cortext command for theassociated footnote;
%% use the ead command for the email address,
%% and the form \ead[url] for the home page:

\title{\textbf{High-efficient Bloch simulation of magnetic resonance imaging sequences based on deep learning}}

\author[inst1]{Haitao Huang\fnref{fn1}}

\affiliation[inst1]{organization={Department of Electronic Science},%Department and Organization
            addressline={Fujian Provincial Key Laboratory of Plasma and Magnetic Resonance, Xiamen University}, 
            city={Xiamen},
            postcode={361005}, 
            country={China}}
\author[inst1]{Qinqin Yang\fnref{fn1}}
\author[inst1]{Jiechao Wang}

\author[inst1]{Pujie Zhang}

\author[inst1]{\\Shuhui Cai}
\author[inst1]{Congbo Cai\corref{cor1}}
\ead{cbcai@xmu.edu.cn}
\cortext[cor1]{Corresponding author}

\fntext[fn1]{Haitao Huang and Qinqin Yang contributed equally to this work.}

\begin{abstract}
%% Text of abstract
\emph{Objective}: Bloch simulation constitutes an essential part of magnetic resonance imaging (MRI) development. However, even with the graphics processing unit (GPU) acceleration, the heavy computational load remains a major challenge, especially in large-scale, high-accuracy simulation scenarios. This work aims to develop a deep learning-based simulator to accelerate Bloch simulation.
\emph{Approach}: The simulator model, called Simu-Net, is based on an end-to-end convolutional neural network and is trained with synthetic data generated by traditional Bloch simulation. It uses dynamic convolution to fuse spatial and physical information with different dimensions and introduces position encoding templates to achieve position-specific labeling and overcome the receptive field limitation of the convolutional network. 
\emph{Main Results}: Compared with mainstream GPU-based MRI simulation software, Simu-Net successfully accelerates simulations by hundreds of times in both traditional and advanced MRI pulse sequences. The accuracy and robustness of the proposed framework were verified qualitatively and quantitatively. Besides, the trained Simu-Net was applied to generate sufficient customized training samples for deep learning-based $\rm T_2$ mapping and comparable results to conventional methods were obtained in the human brain.
\emph{Significance}: As a proof-of-concept work, Simu-Net shows the potential to apply deep learning for rapidly approximating the forward physical process of MRI and may increase the efficiency of Bloch simulation for optimization of MRI pulse sequences and deep learning-based methods.
\end{abstract}


\end{frontmatter}

%% main text
\section{Introduction}
\label{sec:intro}
The non-invasive nature of magnetic resonance imaging (MRI) makes it an outstanding clinical tool in disease diagnosis and research. However, the slow imaging speed and heavy financial burden hinder the development of novel MRI pulse sequences and reconstruction algorithms. Following the underlying physics, MRI simulations allow for rapid iteration of MRI techniques using simple analytical signal expressions or complex numerical modeling. Thus, it becomes an indispensable part of pulse sequence optimization, evaluation and artifact tracing \cite{1drobnjak2006development, 2cai2008sprom,3stocker2010high,4benoit2005simri}. In the era of deep learning (DL), MRI simulation is widely used for training data generation \cite{5yang2022physics,6yang2022model,7frangi2018simulation}, which has spawned a series of novel techniques, including ultra-fast parametric mapping \cite{8cai2018single,9cohen2018mr,10gavazzi2020deep, 11chen2020vivo, 12della2020deepspio}, signal separation \cite{13chen2022ultrafast} and cardiac motion tracking \cite{14loecher2021using}.

One of the main challenges of Bloch simulation is its slow simulation speed. To make the simulation closer to the real situation, many factors need to be considered, such as the shape of the radio frequency (RF) profile, the maximum slew rate of gradient and non-ideal imaging conditions. On the one hand, simulating these factors involves a series of discrete execution intervals, and the computational cost will increase with the complexity of the pulse sequence. On the other hand, two-/three-dimensional Bloch simulations require parallel computation of a large number of evolution-specific spins, which further increases the computation load. To address this, graphics processing unit (GPU) acceleration was introduced and has become the mainstream of complex large-scale MRI simulations \cite{15liu2017fast,16xanthis2013mrisimul,17xanthis2019coremri}. In addition to this, several well-designed software (e.g., SPROM \cite{2cai2008sprom}, MRISIMUL \cite{16xanthis2013mrisimul}, MRiLab \cite{15liu2017fast}) consisting of graphical user interfaces (GUI) were released. The high-performance scientific computing software improves accessibility and allows for more realistic simulation (e.g., multi-pool exchange models) in a personal computer. Despite the great development, traditional MRI simulation is still limited by the computing hardware and has a non-negligible time consumption.

A recent promising and generic approach to speed up simulations is to use machine learning models to approximate the slow forward process. Related applications include but are not limited to modeling molecular energies \cite{18rupp2012fast}, climate science \cite{19kasim2021building}, and even emulating cosmological power spectra \cite{20spurio2022cosmopower}. A powerful tool is physics-informed neural networks (PINN) \cite{21raissi2019physics,22karniadakis2021physics}, which is designed to efficiently solve general nonlinear partial differential equations. By combining the data-driven solution with physical constraints, PINN can represent the continuous solution space within the boundary conditions under the small samples training. With this tool, rapid modeling and prediction of myocardial and cerebral hemodynamics can be achieved in medical imaging \cite{23van2022physics,24sarabian2022physics}. However, among the proposed DL-based simulation methods, only a few works involve MRI and are all limited to accelerate magnetic resonance fingerprinting (MRF) \cite{25balsiger2020learning,26hamilton2019machine,27yang2020game}. An example in Ref \cite{26hamilton2019machine} uses a four-layer neural network for rapid MRF dictionary generation from patient-specific heart rates, which enables online $\rm T_1$ and $\rm T_2$ mapping in the fast-paced clinical workflow. Although the preliminary results are encouraging, the current schemes only consider the evolution of a single voxel, making it difficult to generalize to other MRI techniques, especially pulse sequences with complex spatial encoding \cite{28Cai7837616}.

In this study, a proof-of-concept is presented for the application of deep learning for fast MRI pulse sequence simulation. Although most of the current work is dedicated to using deep learning in solving inverse problems \cite{29Liang,30Agg8434321}, we still try to explore its possibilities in forward physical process approximation. For this purpose, we constructed a two-dimensional fast Bloch simulator based on a convolutional neural network (CNN), called Simu-Net. A dynamic convolution layer \cite{31zhangdodnet,32NIPS2016_8bf1211f} was introduced to encode sequence-specific imaging parameters for physics-modulated simulation. The accuracy of the proposed framework was investigated in one advanced and two classical pulse sequences. The speed of simulation was also compared with that of the mainstream MRI simulation software. Finally, we apply the trained Simu-Net to generate a large number of customized training samples for DL-based $\rm T_2$ mapping and validate it with \textit{in vivo} human brain.

\section{Methods}
\label{sec:methods}

\subsection{Traditional Bloch Simulation}
\label{sec:methods:method1}
To make this paper self-contained, we briefly review the traditional Bloch simulation. Specifically, the spin magnetization vector $\rm \boldsymbol{M} = [M_x, M_y, M_z]^T$ of isochromatic can be expressed by Bloch equation \cite{33bloch1946nuclear}:\begin{small}\begin{equation}\label{eqn-1}
  \begin{aligned}
    {d\boldsymbol{M} \over dt} =   {\gamma} \boldsymbol {M}  \times \boldsymbol{B} - \begin{bmatrix} M_{x}\over T_{2}\\M_{y} \over T_{2}\\M_{z} - M_{0}\over T_{1}\end{bmatrix}
  \end{aligned}
\end{equation}\end{small}where $\gamma$ is the gyromagnetic ratio, $\rm M_0$ is the thermal equilibrium magnetization vector depending on the proton density. The regrowth of the longitudinal magnetization ($\rm M_z$) is characterized by relaxation time constant $\rm T_1$, and the decay of the transverse magnetization ($\rm M_{x,y}$) is governed by constant $\rm T_2$. $\boldsymbol{B}$ denotes the overall field, including the main static field $\rm B_0$ with field inhomogeneity $\Delta{\rm B_0}$, RF field $\boldsymbol{B_1}$ and linear gradient field $\boldsymbol{G}$ at location $\boldsymbol{r}$. It can be modeled in the rotating frame as follows:
\begin{small}
\begin{equation}\label{eqn-2}
  \begin{aligned}
    \boldsymbol{B}(\boldsymbol{r}, t) = (B_{0}  +\Delta{B}_0 (\boldsymbol{r}) +\boldsymbol{G}(t) \cdot\boldsymbol{r} ) \hat{z}+\boldsymbol{B_{1}}(t) 
  \end{aligned}
\end{equation}\end{small}

In general, the acquired MR signals are driven by specific pulse sequences and then reconstructed to images through discrete Fourier transform. Under reasonable assumptions and approximations, the numerical simulation of Bloch equation can be described as successive applications of operators to the magnetization vector, where the initial magnetization vector is given by parametric objects. Despite the complexity of pulse sequences, they can all be represented as linear combinations of three operators, i.e., RF operator, gradient operator and evolution operator. Therefore, transient magnetization vectors $\boldsymbol{M}(\boldsymbol{r},t^+)$/$\boldsymbol{M}(\boldsymbol{r},t^-)$ can be used to record the spin history effect as \cite{2cai2008sprom}:
\begin{footnotesize}
\begin{equation}\label{eqn-3}
	\begin{aligned}
		\boldsymbol{M}(\boldsymbol{r},t^+) &= \boldsymbol{R}_{\alpha} \boldsymbol{M}(\boldsymbol{r}, t^-)
		= \begin{bmatrix}
			1& 0 & 0\\
			0 & cos \alpha & -sin \alpha \\
			0 & sin \alpha & cos \alpha
		\end{bmatrix}
		\begin{bmatrix}
			M_x(\boldsymbol{r}, t^-)\\
			M_y(\boldsymbol{r}, t^-)\\
			M_z(\boldsymbol{r}, t^-)
		\end{bmatrix}\\
		\boldsymbol{M}(\boldsymbol{r},t^+) &= \boldsymbol{G}_{\varphi} \boldsymbol{M}(\boldsymbol{r}, t^-) 
		= \begin{bmatrix}
			cos \varphi & -sin\varphi & 0\\
			sin \varphi & cos \varphi & 0 \\
			0 & 0  & 1
		\end{bmatrix}
		\begin{bmatrix}
			M_x(\boldsymbol{r}, t^-)\\
			M_y(\boldsymbol{r}, t^-)\\
			M_z(\boldsymbol{r}, t^-)
		\end{bmatrix}\\
		\boldsymbol{M}(\boldsymbol{r},t^+) & = \boldsymbol{E}_{T_1,T_2, t} \boldsymbol{M}(\boldsymbol{r}, t^-)
		= \begin{bmatrix}
			e^{-t/T_2} & 0 & 0\\
			0 & e^{-t/T_2} & 0 \\
			0 & 0  & e^{-t/T_1}
		\end{bmatrix}
		\begin{bmatrix}
			M_x(\boldsymbol{r}, t^-)\\
			M_y(\boldsymbol{r}, t^-)\\
			M_z(\boldsymbol{r}, t^-)
		\end{bmatrix}
		+
		\begin{bmatrix}
			0\\
			0\\
			M_z(\boldsymbol{r})^0 (1-e^{-t/T_1})
		\end{bmatrix}
	\end{aligned}
\end{equation}\end{footnotesize}here, for RF operator $\boldsymbol{R}_\alpha$, a simple rotation operation with flip angle $\alpha$ is applied to represent the effect of RF pulse, and the transmitting RF coil inhomogeneity ($\rm B_1^+$) impacts the strength of the flip angle. The gradient operator $\boldsymbol{G}_{\varphi}$ is the time-varying linear gradient field for spatial encoding, where $\varphi = \gamma \boldsymbol{G}(\boldsymbol{r},t)\cdot(t^+-t^-)$. Exponential decay terms of $\rm T_1$ and $\rm T_2$ relaxations (respectively, $1-e^{-t/T_1}$ and $e^{-t/T_2}$) are performed using the operator $\boldsymbol{E}_{T_1,T_2,t}$. The effects of susceptibility-induced $\rm B_0$ inhomogeneities are present throughout the pulse sequence, resulting in additional phase accumulation in spins. At the signal acquisition stage, the integrals of all magnetization vectors are assigned to each voxel in k-space according to the desired sampling trajectory. 

\subsection{Bloch Simu-Net}
\label{sec:methods:blochsimunet}
A deep learning framework with non-linear activation units, termed Bloch Simu-Net, is used to approximate the forward physical process of Bloch simulation. A schematic demonstration of the proposed framework is shown in Figure 1. Briefly, parametric templates ($\rm T_1$, $\rm T_2$, $\rm M_0$, $\rm B_0$ and $\rm B_1$), $\textit{P}_\textit{j}$, and position encoding template (PET), $\textit{P}_\textit{e}$, were first created to initialize the spatial distribution of various properties of virtual objects. The scan-specific pulse sequence parameters, $\textit{P}_\textit{s}$, were used to generate dynamic convolution kernels through a multilayer perceptron, $\mathcal{F}$, for sequence-informed encoding. Finally, a simulated MR image, $\textit{I}_\textit{w}$, can be obtained from an encoder-decoder network, $\mathcal{N}$, by feeding with encoded feature maps, $\textit{f}$. Thus, the nonlinear forward physical process can be simplified as:

\begin{equation}\label{eqn-4}
  \begin{aligned}
    \textit{I}_w = \mathcal{N}(f_{\mathcal{F}(\boldsymbol{P}; \theta_{\mathcal{F}})}; \theta_{\mathcal{N}})
  \end{aligned}
\end{equation}where, $\boldsymbol{P}$ denotes the combination of $\textit{P}_\textit{j}$, $\textit{P}_\textit{e}$ and $\textit{P}_\textit{s}$.  $\theta_\mathcal{N}$ and $\theta_\mathcal{F}$ are the trainable parameters of the encoder-decoder network and the multilayer perceptron, respectively. In practice, the encoder-decoder network is a five-level U-Net \cite{34falk2019u}. The details of parametric templates, PET and dynamic convolution module are described in the following sections.

\begin{figure}[!htb]
\centering
\includegraphics[width=\textwidth]{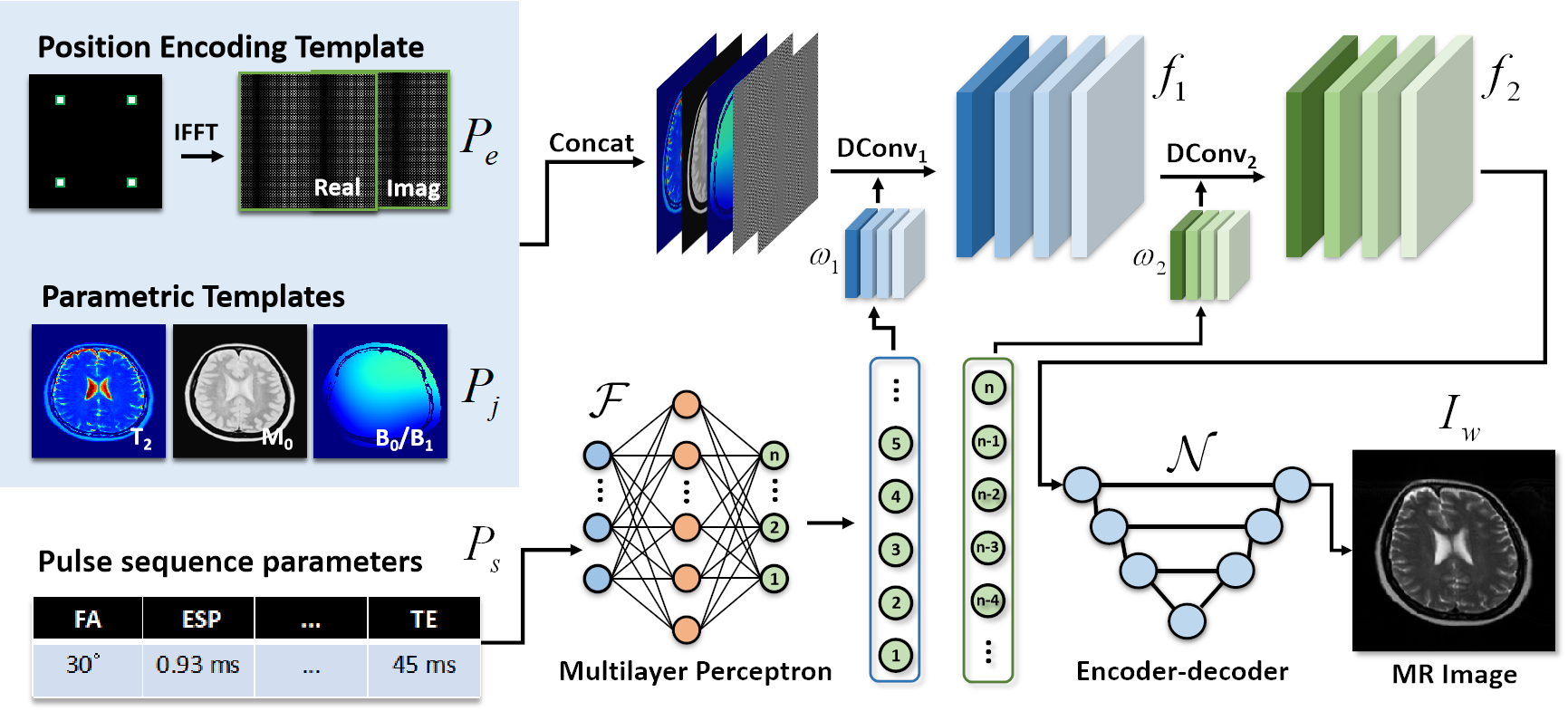}
\captionsetup{labelfont=bf}
\caption{Schematic illustration of Simu-Net for fast simulation of MRI sequences. The complex-valued position encoding template and parametric templates ($\rm T_1$, $\rm T_2$, $\rm M_0$, $\rm B_0$ and $\rm B_1$) were concatenated and encoded to sequence-informed feature maps $f$ by dynamic convolution. The weights and bias of dynamic convolution were generated by pulse sequence parameters and learnable multi-layer perceptron. The encoder-decoder is an U-Net.}
\label{fig1}
\end{figure}

\subsubsection{Parametric Templates}
\label{sec:methods:blochsimu:PT}
For 2D spatial encoding in Bloch simulations, the spatial distribution of tissue parameters ($\rm T_1$, $\rm T_2$, $\rm M_0$) and field inhomogeneity ($\Delta{\rm B_0}$, $\rm B_1$) should be constructed as parametric templates. In this work, the tissue parameters were synthesized using multi-contrast MRI data from the IXI database (\url{http://brain-development.org/ixi-dataset/}) following the MOST-DL method \cite{6yang2022model}. In brief, the registered $\rm PD$-weighted and $\rm T_2$-weighted images were firstly transformed to pseudo parametric maps and then upsampled to $512 \times 512$ grids, in which $\rm T_2 \in [0, 650]$ ms, $\rm M_0 \in [0, 1]$. The field inhomogeneity can be modeled as a random surface through 2D low-order polynomial functions and Gaussian functions. Finally, the $\Delta{\rm B_0}$ field is bounded within $\pm 150$ Hz and the $\rm B_1^+$ field is scaled to the range of $0.7 \thicksim 1.3$.

\subsubsection{Position Encoding Template}
\label{sec:methods:blochsimu:PET}
Some advanced MRI sequences, such as the spatiotemporal encoding imaging method \cite{35schmidt2014new} or MOLED acquisition \cite{8cai2018single, 36zhang2019robust}, involve special spatial encoding, which results in nonlocal signal modulation patterns in the image domain. However, the translation invariance nature of CNN makes it impossible to accurately learn the mapping relationship between parametric templates and MRI signals at different spatial positions. Inspired by positional encoding in Transformer\cite{vaswani2017attention}, we introduce PET to label different spatial positions of parametric templates and overcome the receptive field limitation of convolutional networks. Specifically, a zero-filled complex matrix of the same size as the target image is firstly created, then $1+1j$ is assigned to the symmetric position of the four quadrants. After the inverse Fourier transform, a complex-valued PET is obtained. Before being fed into the network, the PET was divided into real and imaginary components as two individual channels. A representative example of PET is shown in Supporting Information Figure S1, where there is a high-frequency periodic signal modulation along the horizontal and vertical directions.

\subsubsection{Sequence-informed Dynamic Convolution}
\label{sec:methods:blochsimu:SDC}
How to encode one-dimensional sequence information into network training is the main challenge of Simu-Net. Inspired by dynamic filter learning \cite{32NIPS2016_8bf1211f}, a dynamic filter generation module was introduced to generate the dynamic convolution kernels by feeding pulse sequence parameters (e.g., flip angles, echo spacing, echo time), as shown in Figure 1. The subsequent encoder-decoder network will be uncertain if the series of varying imaging parameters are unknown. Hence, the dynamic convolutional module should be placed before the encoder-decoder network to fusion the inputs and sequence parameters, termed sequence-informed dynamic convolution. Specifically, a 3-layer perceptron is used as a non-linear encoder $\mathcal{F}(\cdot)$ with trainable parameters $\theta_\mathcal{F}$. The $P_{s}$ is the vector consisting of a set of normalized sequence parameters as the input of $\mathcal{F}(\cdot)$. Then, the weights and bias of dynamic convolution kernels, $W = [\omega_1 ,..., \omega_i]$, can be obtained as follows:

\begin{equation}\label{eqn-5}
  \begin{aligned}
    W = [\omega_1 ,..., \omega_i] & = \mathcal{F}(P_{s};\theta_F)
  \end{aligned}
\end{equation}

\noindent where $i$ denotes the number of dynamic convolutional layers. In this work, two dynamic convolutional layers (i.e., $i = [1,2]$) with $1 \times 1$ kernel were constructed from the dynamic weights and were applied to combined parametric templates and PET as:

\begin{equation}\label{eqn-6}
  \begin{aligned}
    f_1 & = DConv_1( concat(P_e, P_j); \omega_1)\\
    f_2 & = DConv_2( ReLU(f_1); \omega_2)
  \end{aligned}
\end{equation}

\noindent where each dynamic convolution contains 16 filters. Finally, a total of 16 parameters-encoded feature maps were obtained as the inputs of the further encoder-decoder network. The number of weights and biases of the dynamic convolution layers determines the output size of the multi-layer perceptron. In this work, the kernel size of dynamic convolution is $1\times1$, and the output size of the multi-layer perceptron is $n = 368$, that is, there are 96 parameters ($5\times16$ weights and 16 biases) in the first dynamic convolutional layer and 272 parameters ($16\times16$ weights and 16 biases) in the second layer. The number of nodes in the hidden layer is $4 \times n = 1472$ for expanding the feature space.

\subsection{Experiments}
\label{sec:methods:simuexp}
The traditional Bloch simulations were performed on MRiLab \cite{15liu2017fast} (\url{ https://leoliuf.github.io/MRiLab/ }) and SPROM \cite{2cai2008sprom} (\url{https://doi.org/10.6084/m9.figshare.19754836.v2}) software. Three different MRI pulse sequences were selected as examples to validate our proposed method, including fast spin echo (FSE), gradient-echo echo-planar imaging (GRE-EPI) and multiple overlapping-echo detachment planar imaging (MOLED). Figure 2 gives the sequence diagrams of these three pulse sequences. In this work, we only considered the following parameters that most govern the signal: echo time (TE), echo spacing (ESP), echo train length (ETL), flip angles of excitation (FA) and refocusing RF pulse (Re-FA) and echo-shifting gradients (SG). Detailed variable imaging parameters considered in Simu-Net training are listed in Table 1. The spatial resolution of all experiments is fixed
at $1.7\times1.7$ mm$^2$ ($\rm FOV = 220\times220$ mm$^2$, matrix size = $128\times128$), and the magnetization is set to recover instantaneously after signal sampling (i.e., the repetition time, $\rm TR$, can be ignored).

\begin{figure}[!htb]
\centering
\includegraphics[width=\textwidth]{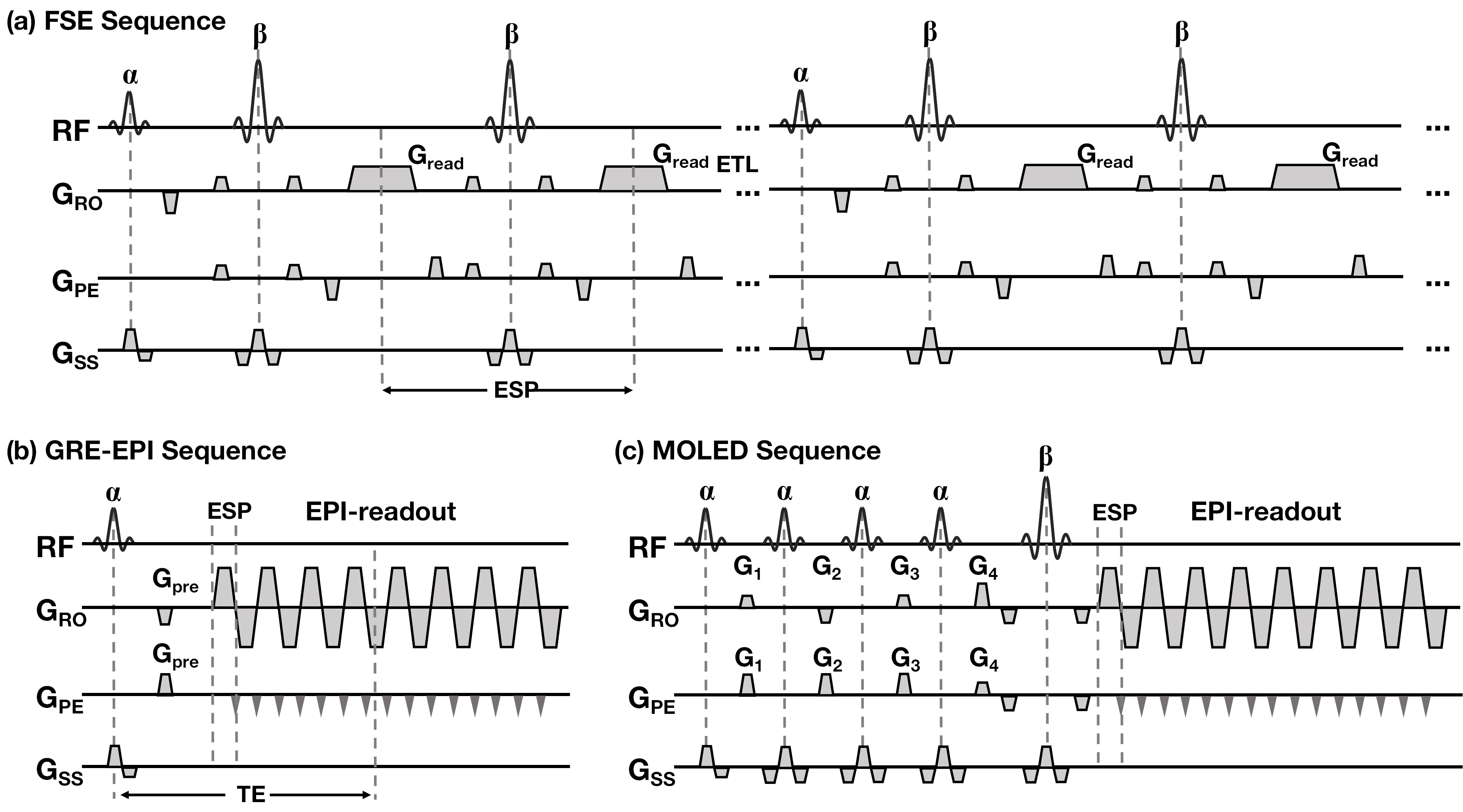}
\captionsetup{labelfont=bf}
\caption{The sequence diagrams of (a) fast spin echo (FSE), (b) gradient-echo echo planar imaging (GRE-EPI) and (c) multiple overlapping-echo detachment (MOLED). In MOLED, four excitation pulses and a refocusing pulse are used to prepare spin echoes with different echo times followed by an EPI readout module. The positions of different echoes in k-space are determined by four echo-shifting gradients ($\rm G_1$, $\rm G_2$, $\rm G_3$ and $\rm G_4$). ESP: echo spacing; ETL: echo train length; EPI: echo planar imaging. }
\label{fig2}
\end{figure}

\subsubsection{Fast Spin Echo Simulation}
\label{sec:methods:simuexp::FSE}
FSE sequence has been commonly used in clinical practice to obtain high-resolution $\rm T_2$-weighted MR images \cite{37bernstein2004handbook}. Figure 2(a) illustrates a typical multi-shot FSE sequence, where one shot is composed of one excitation RF pulse followed by multiple refocusing pulses. In this sequence, $\rm B_1^+$ field inhomogeneity, ESP and FA are considered to significantly affect the contrast of final MR images; thus, these parameters were randomly sampled from the empirical ranges (Table 1). Besides, the fixed parameters are as follows: shot number = 8, ETL = 16, Re-FA = $180^{\circ}$. The TE of the FSE sequence is governed by the varying ESP.

\subsubsection{Gradient-echo Echo-planar Imaging Simulation}
\label{sec:methods:simuexp::GRE-EPI}
Single-shot GRE-EPI sequence plays a vital role in BOLD functional MRI \cite{43biswal1995functional} and perfusion MRI \cite{44edelman1994qualitative}. However, its sensitivity to $\rm B_0$ field inhomogeneity can distort the resulting images. A typical GRE-EPI sequence is shown in Figure 2(b). $\rm G_{pre}$ represents the prephase gradient, which is typically set to $1/2$ the area of phase-encoding gradients to ensure that the echo is centered in k-space. For GRE-EPI simulation, TE and ESP were selected in dynamic filter generation since those parameters affect the distortion, and FA was fixed to $90^{\circ}$.

\begin{table}[!htb]
	\begin{center}
	\captionsetup{labelfont=bf}
	\caption{Parametric templates, non-ideal conditions, imaging parameters for three MRI sequences. Re-FA is the flip angle of refocusing pulse. SG is the applied shift gradient.}
	\resizebox{\textwidth}{!}{
		\begin{tabular}{c|cc|cc|cccccc}
		% \begin{tabular}{ppppppppppp}  
		\hline
		 \multirow{1}{2em}{}  & \multicolumn{2}{c|}{\multirow{1}{5em}{Parametric Template}} & \multicolumn{2}{c|}{\multirow{1}{5em}{Non-ideal Condition}} & \multicolumn{6}{c}{\multirow{1}{6em}{Imaging Parameter}}  \\
		 &\multicolumn{2}{c|}{}&\multicolumn{2}{c|}{}& \multicolumn{6}{c}{}
		 \\
		\hline
		 & $\rm T_2$ & $\rm M_0$ & $\Delta{\rm B_0}$ & $\rm B_{1}^+$ & TE & ESP & ETL & FA & Re-FA & SG \\ \hline
		FSE & $\checkmark$ & $\checkmark$ & $\backslash$ & $0.7 \thicksim 1.3$ & $\backslash$ & $8\thicksim15$ ms & 16 & $20\thicksim70^{\circ}$ & $180^{\circ}$ & $\backslash$ \\
		GRE-EPI & $\checkmark$  & $\checkmark$  & $-150\thicksim+150$ Hz  & $\backslash$ & 40-80 ms & $0.375\thicksim0.55$ ms & $\backslash$ & $90^{\circ}$ & $\backslash$ & $\backslash$ \\
		MOLED & $\checkmark$ & $\checkmark$ & $\backslash$ & $0.7 \thicksim 1.3$ & $\backslash$ & $0.375\thicksim0.54$ ms & $\backslash$ &$30^{\circ}$ & $180^{\circ}$ & $\pm{5}\%$ \\
		\hline
		\end{tabular}
	}
	
\end{center}
\end{table}

\subsubsection{Multiple Overlapping-echo Imaging Simulation and $\rm T_2$ mapping}
\label{sec:methods:simuexp::MOLED}
Single-shot MOLED sequence is a recently proposed technique for ultra-fast $\rm T_2$ mapping \cite{36zhang2019robust}. As shown in Figure 2(c), MOLED uses four excitation pulses and a refocusing pulse to prepare multiple spin echoes with different TEs. The echo-shifting gradients $\rm G_1-G_4$ shift the position of spin echoes in k-space. Through the EPI readout module, a whole image containing multiple echoes can be obtained within a few hundred milliseconds. In this sequence, ESP and SG will affect the range of TEs, FA = $30^{\circ}$ and Re-FA = $180^{\circ}$.

 For MOLED $\rm T_2$ mapping, we followed the workflow of MOST-DL \cite{6yang2022model}, where the Bloch simulation was replaced by the proposed Simu-Net for generating training samples. A five-level U-Net \cite{34falk2019u} was trained by synthetic data to learn the end-to-end mapping from complex-valued MOLED images to $\rm T_2$ maps. During the training and testing phase, the $128\times128$ overlapping-echo image was zero-padded to $256\times256$ in k-space and then normalized in the image domain.

For validating the Simu-Net on synthetic data generation for MOLED $\rm T_2$ mapping, \textit{in vivo} MOLED data were acquired from one healthy volunteer on a whole-body MRI system at 3T (MAGNETOM Prisma TIM, Siemens Healthcare, Erlangen, Germany) with a 20-channel head coil. The study protocol was approved by the institutional research ethics committees, and written informed consent was obtained from the volunteer. The imaging parameters were as follows: TE = 22.0, 52.0, 82.0, 110.0 ms, TR = 8000 ms, FOV = $220\times220$ mm$^2$, matrix size = $128\times128$, GRAPPA = 2, FA = $30^{\circ}$, slice number = 21, slice thickness = 4.0 mm. For comparison, we collected MOLED data with three different ESP (i.e., 1.08, 0.93 and 0.75 ms). Noted that the ESP of the readout gradient is 2 times that of the simulation to maintain the consistency of the ETL. The reference $\rm T_2$ maps were obtained from the SE sequence using the TEs = 35, 50, 70, 90 ms, TR = 2500 ms. The total scan times are 8.0 s for the MOLED and 21 min for reference SE sequence.

\subsection{Implementation and validation}
\label{sec:methods:IandV}
In this work, three pulse sequences are treated as different non-linear functions. Thus, we trained the neural networks separately for different pulse sequences. We use $512\times512$ parametric templates to generate $128\times128$ images through Bloch simulation with SPROM software \cite{2cai2008sprom}. Then, the original parametric templates were down-sampled to $128\times128$ as the network inputs. The sample size of training set is 2000 for each task, and additional 100 samples were generated for network evaluation.  The parameter range of the test samples is the same as that of the training samples (as mentioned in Table 1), but the parametric templates (i.e., the brain tissues) used for testing are beyond the training samples. The learning rate was initially set as $1.2\times 10^{-4}$ and decayed every 25 epochs with $80\%$. $\rm L_1$ norm was chosen as the loss function and the total number of epochs is 300 for all experiments. The training times of FSE, GRE-EPI and MOLED are 2 h 30 min, 7 h 50 min and 5 h 12 min, respectively (Supporting Information Figure S2 provides the loss curves). The network training and testing pipelines were coded using PyTorch library. All processes were performed on a machine with an NVIDIA 2080Ti GPU. 

To quantitatively evaluate the results, peak signal-to-noise ratio (PSNR) and structural similarity index (SSIM) were calculated between simulated results of the Simu-Net and Bloch equation. In addition, we qualitatively and quantitatively evaluated the results of MOLED $\rm T_2$ mapping with different imaging parameters, where linear regression analysis was performed on the \textit{in vivo} experiments at the region-of-interest (ROI) level. These ROIs were manually selected from gray/white matter and then were used to calculate the mean $\rm T_2$ values.

\section{RESULTS}
\label{sec:results}
The results of Bloch equation and deep learning-based Simu-Net for FSE, GRE-EPI and MOLED are shown in Figure 3. For FSE and GRE-EPI, the quality of simulated images from Simu-Net agrees well with those from Bloch simulation, even with the effect of $\rm B_1$/$\rm B_0$ inhomogeneity (Figure 3(a) and 3(b)). Due to the low acquisition bandwidth along the phase-encoding direction of GRE-EPI, $\rm B_0$ inhomogeneity causes large phase accumulation and leads to severe geometric distortions, making the GRE-EPI more difficult to learn. Thus, the errors maps show the precision of FSE (PSNR = 53.23 dB, SSIM = 0.998) simulation is higher than that of distortion-corrupted GRE-EPI (PSNR = 36.75 dB, SSIM = 0.937). The results of distortion-free GRE-EPI are provided in Supporting Information Figure S3.  As for MOLED, high-fidelity magnitude (PSNR = 37.65 dB, SSIM = 0.972) and phase can be simultaneously obtained by Simu-Net, as shown in Figure 3(c). The zoomed-in views demonstrate that the Simu-Net can achieve high-accuracy simulation even with complex signal modulation.

\begin{figure}[!htb]
\centering
\includegraphics[width=\textwidth]{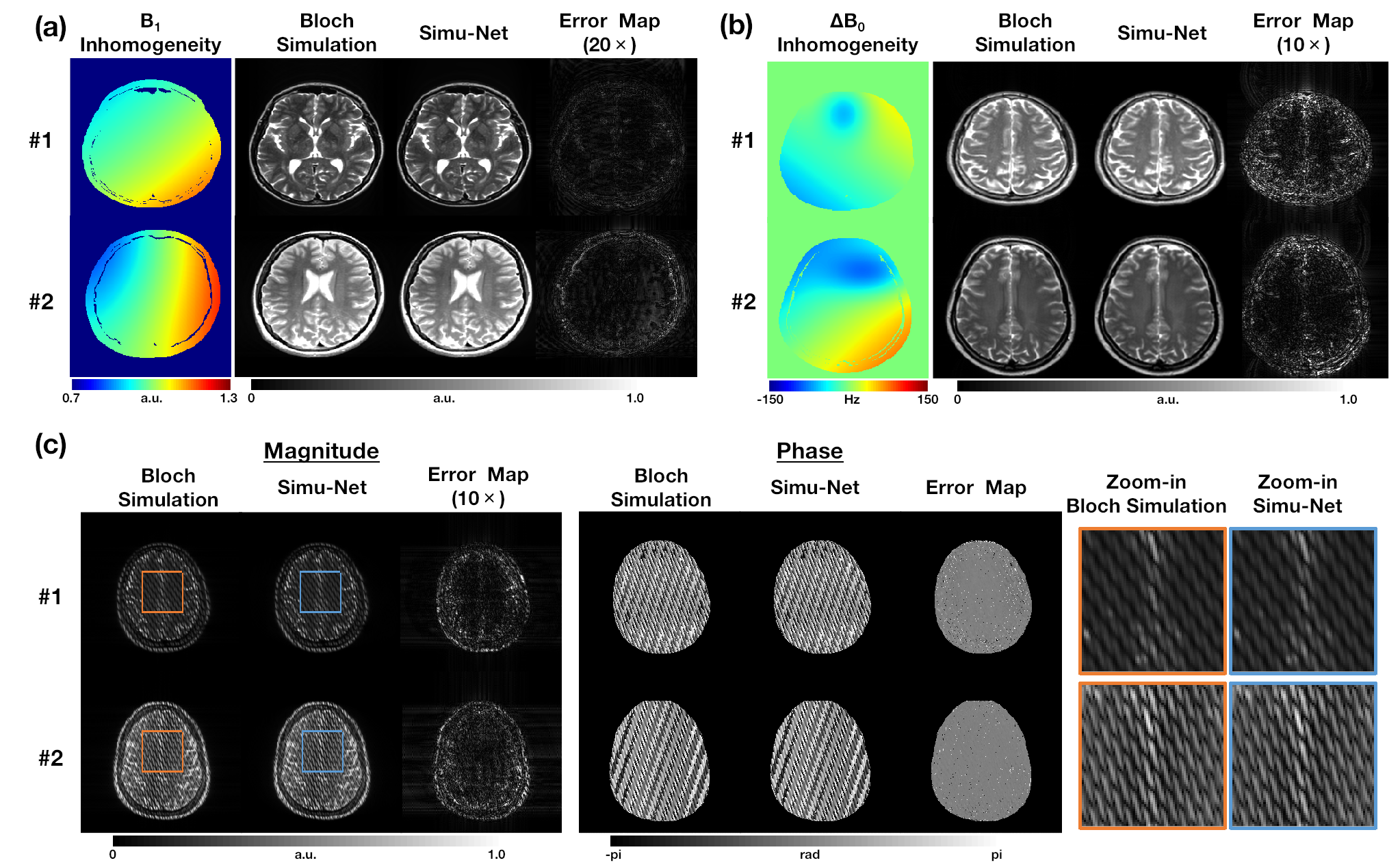}
\captionsetup{labelfont=bf}
\caption{The simulation results of Bloch equation and deep learning-based Simu-Net. (a) The results of FSE simulation with $\rm B_1$ inhomogeneity, case 1: FA = $55.72^{\circ}$, ESP = 12.0 ms; case 2: FA = $56.16^{\circ}$, ESP = 11.6 ms. (b) The results of GRE-EPI simulation with $\Delta{\rm B_0}$ inhomogeneity, case 1: TE = 45.6 ms, ESP = 0.544 ms; case 2: TE = 51.2 ms, ESP = 0.446 ms. (c) The results of MOLED simulation, case 1: ESP = 0.54 ms; case 2: ESP = 0.375 ms.}
\label{fig3}
\end{figure}

Table 2 compares the speed of MRI simulation between Bloch simulation and Simu-Net. Two mainstream software (MRiLab and SPROM) were used to perform Bloch simulations under GPU acceleration. It can be seen that the time required to simulate an MRI image is specific to the pulse sequence, where the GRE-EPI takes about 12 seconds, MOLED about 30 seconds, and FSE about 500 or 1300 seconds. Surprisingly, Simu-Net enables millisecond-level simulation of different pulse sequences and increases the simulation speed of FSE sequences by 8728 times.

\begin{table}[h!]
\begin{center}
\captionsetup{labelfont=bf}
\caption{The comparison of speed for synthesizing one data.}
\begin{tabular}{ c c c c } 
\hline
Sequence & MRiLab & SPROM & Simu-Net \\
\hline
 FSE & 497.53 s & 1320.20 s & 0.057 s\\ 
 GRE-EPI & 11.44 s & 29.81 s & 0.064 s \\ 
 MOLED & 12.16 s & 32.42 s & 0.073 s \\ 
 \hline
\end{tabular}

\end{center}
\end{table}

Figure 4 shows the results of GRE-EPI simulation from Bloch equation and Simu-Net for investigating the $\rm B_0$-introduced image distortion and TE-dependent signal attenuation. The spatial distribution of the $\Delta{\rm B_0}$ field used in simulations is shown in Figure 4(a). In Figure 4(b), GRE-EPI images with different geometric distortions can be obtained by Simu-Net at an increased $\Delta{\rm B_0}$ field, even the severe distortions (white arrows)
due to $\pm150$ Hz inhomogeneous field. In addition, the simulated signal attenuation of Simu-Net from 5 ROIs (marked by red) agrees well with that of Bloch simulation, where the signal intensity decreases with increasing echo time (Figure 4(c)).
\begin{figure}[!htb]
\centering
\includegraphics[width=\textwidth]{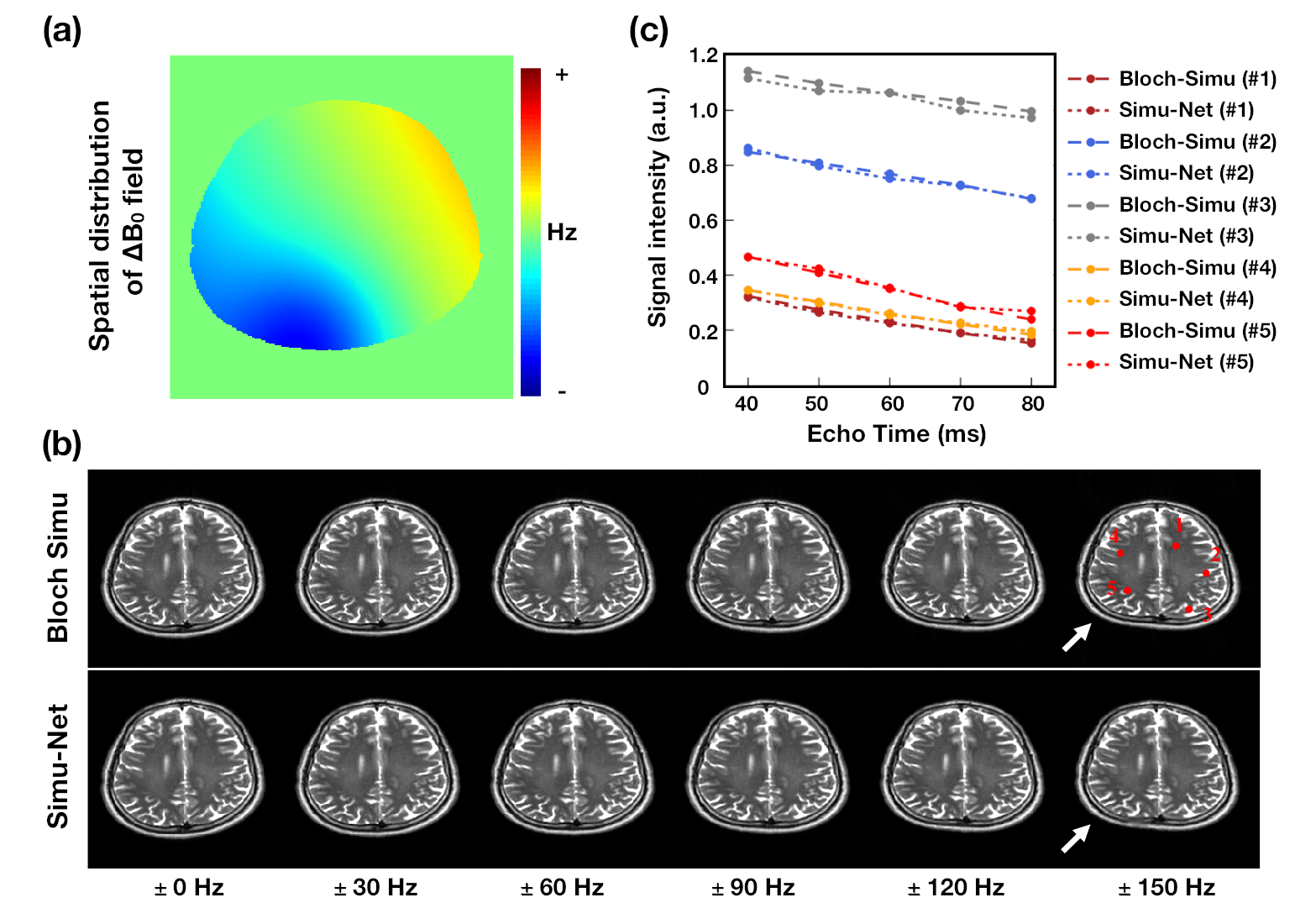}
\captionsetup{labelfont=bf}
\caption{The results of GRE-EPI simulation from Bloch equation and Simu-Net. The $\Delta{\rm B_0}$ with the same spatial distribution (a) but different intensities were used in Bloch simulation and Simu-Net. (b) The line chart. (c) shows the changes of signal intensity over echo times from 5 ROIs on the $\pm150$ Hz case (marked by red).}
\label{fig4}
\end{figure}

Figure 5 examines the effects of spin numbers on the MOLED simulation. Here, all experiments take into account the effect of $\rm B_1$ inhomogeneity and use the same imaging parameters. For traditional Bloch simulation, the insufficient number of spins (e.g., $256\times256$ or $128\times128$) leads to signal corruption in the final $128\times128$ MOLED images, and additional echoes can also be observed in k-space (marked by white arrows). In contrast, Simu-Net achieves high-accuracy simulations with a limited number of spins, delivering high-fidelity strip patterns in the image domain and almost lossless signal quality in k-space.

\begin{figure}[!htb]
\centering
\includegraphics[width=\textwidth]{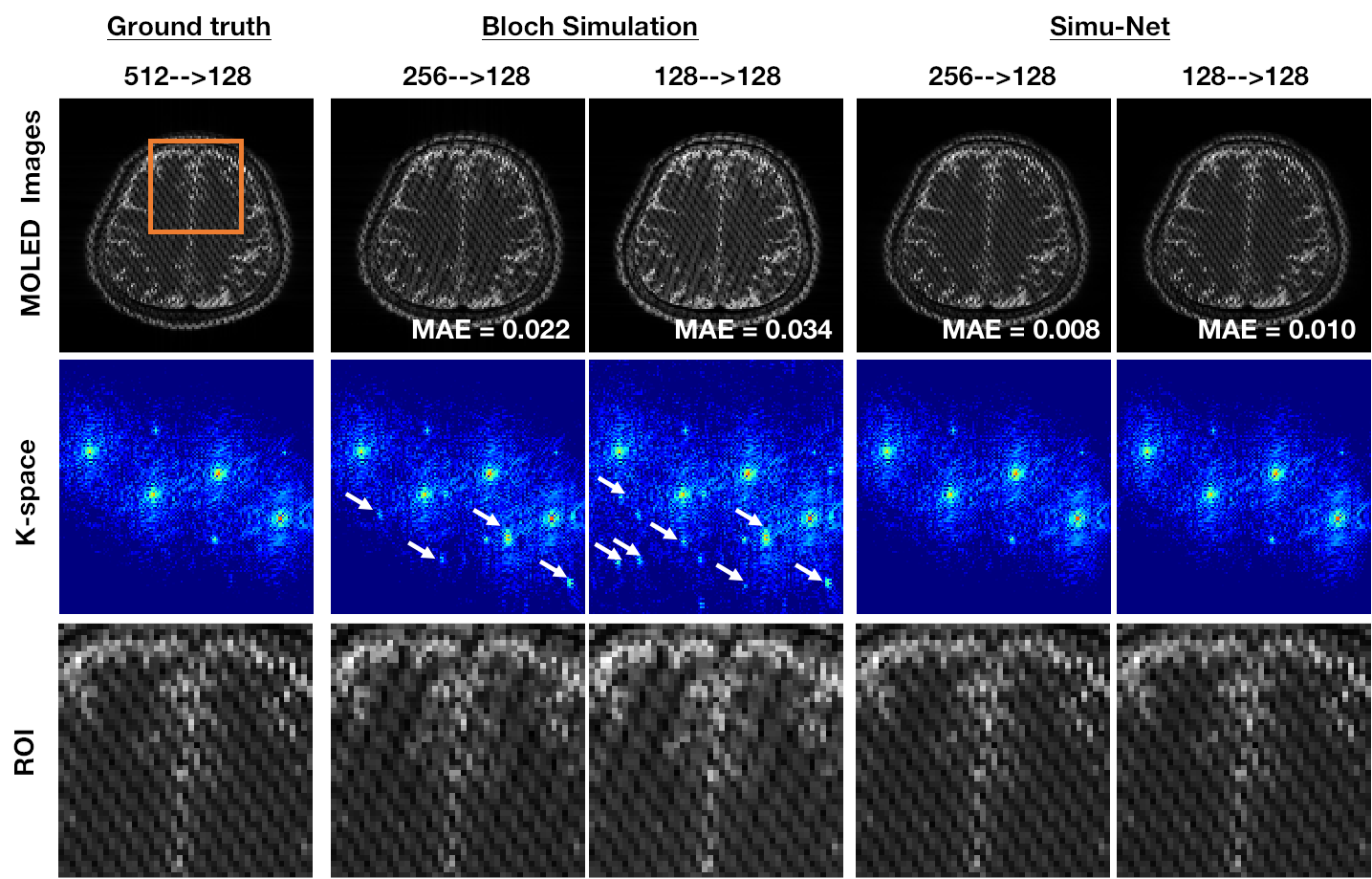}
\captionsetup{labelfont=bf}
\caption{The effects of template size (the number of spins) on MOLED simulation. The insufficient number of spins ($256\times256$ or $128\times128$) result in image degradation and additional echoes in k-space (white arrows) for Bloch simulation. For Simu-Net, high-accuracy MOLED simulation can be achieved by using templates with a relatively small size.}
\label{fig5}
\end{figure}

Figure 6 shows the \textit{in vivo} results of MOLED $\rm T_2$ mapping, where the training samples were generated by Bloch simulation and Simu-Net. To reconstruct MOLED images with different acquisition parameters (ESP = 1.08, 0.93 and 0.75 ms), three independent training sets (4000 samples each) were obtained, which took 40 hours for Bloch simulation and 15 minutes for Simu-Net. It can be seen that Simu-Net enables MOLED $\rm T_2$ mapping with image quality close to that of Bloch simulation and significantly reduces the time of data generation.

\begin{figure}[!htb]
\centering
\includegraphics[width=\textwidth]{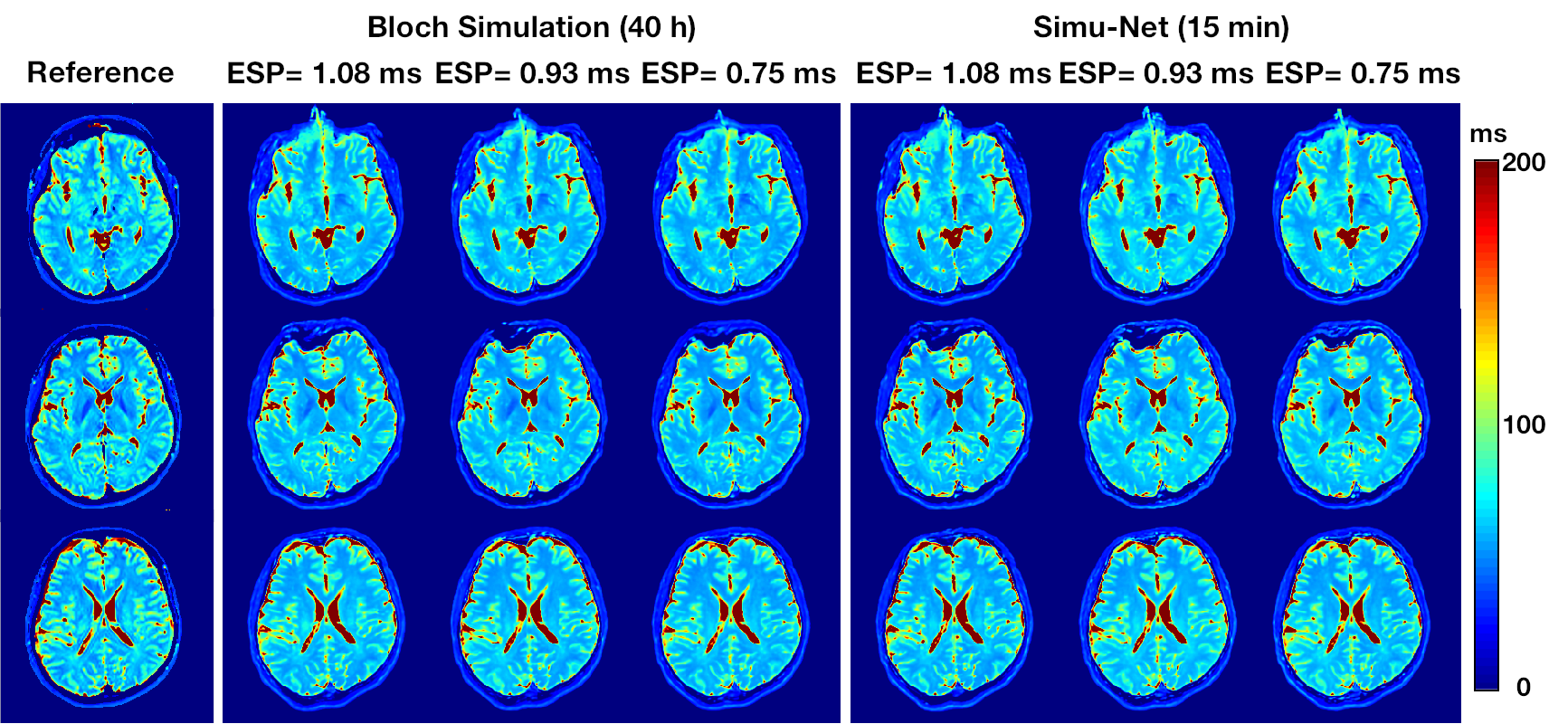}
\captionsetup{labelfont=bf}
\caption{Comparison of MOLED $\rm T_2$ mapping for \textit{in vivo} human brain with different acquisition parameters (ESP = 1.08, 0.93 and 0.75 ms) from Bloch simulation samples and Simu-Net samples. The reference $\rm T_2$ maps were obtained by SE sequence with TE = 35, 50, 70, 90 ms. It took 40 hours for Bloch simulation to generate 12000 training samples and about 15 minutes for Simu-Net.}
\label{fig6}
\end{figure}

Linear regression plots quantitatively compare the $\rm T_2$ values derived from the Bloch simulation and Simu-Net, as shown in Figure 7. A total of 72 ROIs covering the whole brain were manually selected from grey matter and white matter to calculate the mean $\rm T_2$ values. The quantitative values of both methods in these ROIs are relatively consistent with the reference method, with $\rm R^2$ greater than 0.9. However, the results of Simu-Net for different imaging parameters are slightly less consistent than those of Bloch simulation, with $\rm R^2$ decreasing from 0.947, 0.933, and 0.942 to 0.934, 0.924, and 0.919, respectively.

\begin{figure}[!htb]
\centering
\includegraphics[width=\textwidth]{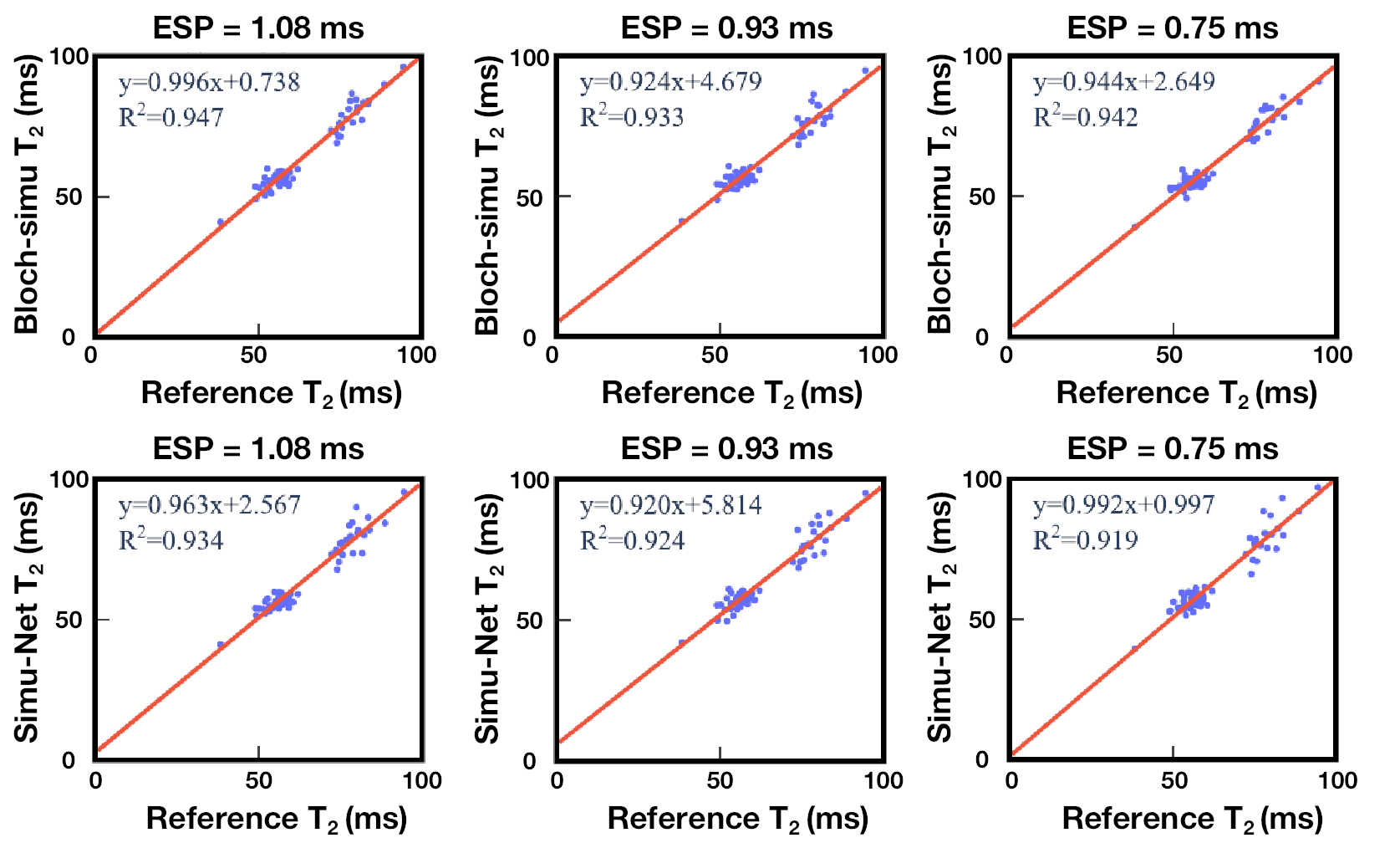}
\captionsetup{labelfont=bf}
\caption{Linear regression analysis of $\rm T_2$ maps obtained by using MOLED and reference method in different echo spacing (1.08, 0.93 and 0.75 ms). The training samples were generated by Bloch simulation (top row) and Simu-Net (bottom row), respectively.}
\label{fig7}
\end{figure}

\section{DISCUSSION}
\label{sec:discussion}

In this work, we proposed a deep learning-based framework, Simu-Net, for ultra-fast MRI simulation. To the best of our knowledge, Simu-Net is the first method for approximating the 2D Bloch equation using convolutional neural networks. As the core component of this framework, sequence-informed dynamic convolution is designed to encode scan-specific parameters to generate physically constrained MRI images. The framework was demonstrated in an advanced and two classical pulse sequences, where Simu-Net was shown to efficiently and accurately simulate complex-valued signals even with field inhomogeneity. The application of Simu-Net for fast training data generation was also demonstrated through MOLED simulation for \textit{in vivo} $\rm T_2$ mapping under various imaging parameters. The preliminary results show that Simu-Net can significantly facilitate the application of synthetic-data-driven deep learning and achieve MRI simulations close to Bloch equation with a speedup of hundreds of times.

Dynamic convolution was firstly proposed by Jia et al. to improve the performance of ImageNet classification by introducing attention mechanisms \cite{32NIPS2016_8bf1211f} without increasing the depth and width of the network. However, in Simu-Net, dynamic convolution was used to transform pulse sequence parameters into learnable convolution kernels and enable simultaneous encoding of 2D (parametric templates) and 1D (imaging parameters) information to high-dimensional feature maps. This strategy improves the flexibility of physics-informed network training rather than simple end-to-end mapping. On the other hand, Simu-Net is trained by a few randomly sampled synthetic data similar to classical PINN. Although these data are discrete points in a high-dimensional solution space, Simu-Net achieves continuous data-driven solutions to Bloch equations. From the results of Figure 4, we see that TE-dependent signal attenuation and $\rm B_0$-introduced geometric distortion of GRE-EPI can be successfully simulated, closing to Bloch equations.

Due to the fast forward inference of convolutional neural networks, Simu-Net can significantly speed up Bloch simulations and surpass the GPU-accelerated high-performance scientific computing software. The results in Table 2 show that accelerations of more than 8000 times can be achieved for FSE simulation. Compared with GRE-EPI and MOLED, FSE with hundreds of pulses needs to simulate much more discrete spin evolution at a small-time interval, dramatically increases the required computational time. In contrast, Simu-Net performs MRI simulations in the form of domain transformation, ignoring the spin history effects and avoiding complex signal encoding. Based on this, synthetic-data-driven deep learning may no longer be limited by the computational burden of data generation. Although Simu-Net also uses synthetic data for network training and requires retraining for new downstream tasks, it is essentially a function approximation of the forward physical process and thus needs a much smaller sample size and shorter training time than the inverse problem. Figure 8 shows the effects of different sample sizes on MRI simulation for Simu-Net. It can be seen that less than 2000 samples is enough to train Simu-Net for high-accuracy MRI simulation, even for MOLED with complex spatial encoding.

In Simu-Net, CNN priors and PET are believed to jointly contribute to high-accuracy MOLED simulation. Unlike FSE and GRE-EPI, multiple echoes prepared by independent RF pulses result in strip patterns in the image domain of MOLED signal, which means that the signal is modulated by a special point spread function. To adapt the position-specific signal modulation, PET was constructed with high-frequency elements and concatenated with parametric templates as network input. Our experiments show that the training of MOLED simulation is difficult to converge without the input of position encoding. Although classical sequences are not affected by this, the introduction of position encoding is still considered as a generalized extension of Simu-Net.

The experiment in Figure 5 is based on a non-negligible issue, that is, there is a trade-off between simulation accuracy and simulation time. Traditional Bloch simulation involves discrete spin evolution and gradient encoding,  which require the spatial discretization grid to exceed the fastest gradient spiral by at least a factor of two (i.e., spatial Nyquist condition \cite{45nyquist}). Sub-Nyquist sampling results in catastrophic loss of accuracy, especially for FSE or MOLED with multiple RF pulses or complex spatial frequency encoding. In contrast, Simu-Net performs simulation through image-to-image transformation and uses high-accuracy simulation results as labels. A series of non-linear operations is replaced by a deep neural network, avoiding direct Fourier encoding that enables simulation to ignore the traditional sampling limitations. Thus, Simu-Net improves spatial and temporal efficiency of MRI simulation and may be used for large-scale simulation in the future.
\begin{figure}[!htb]
	\centering
	\includegraphics[width=\textwidth]{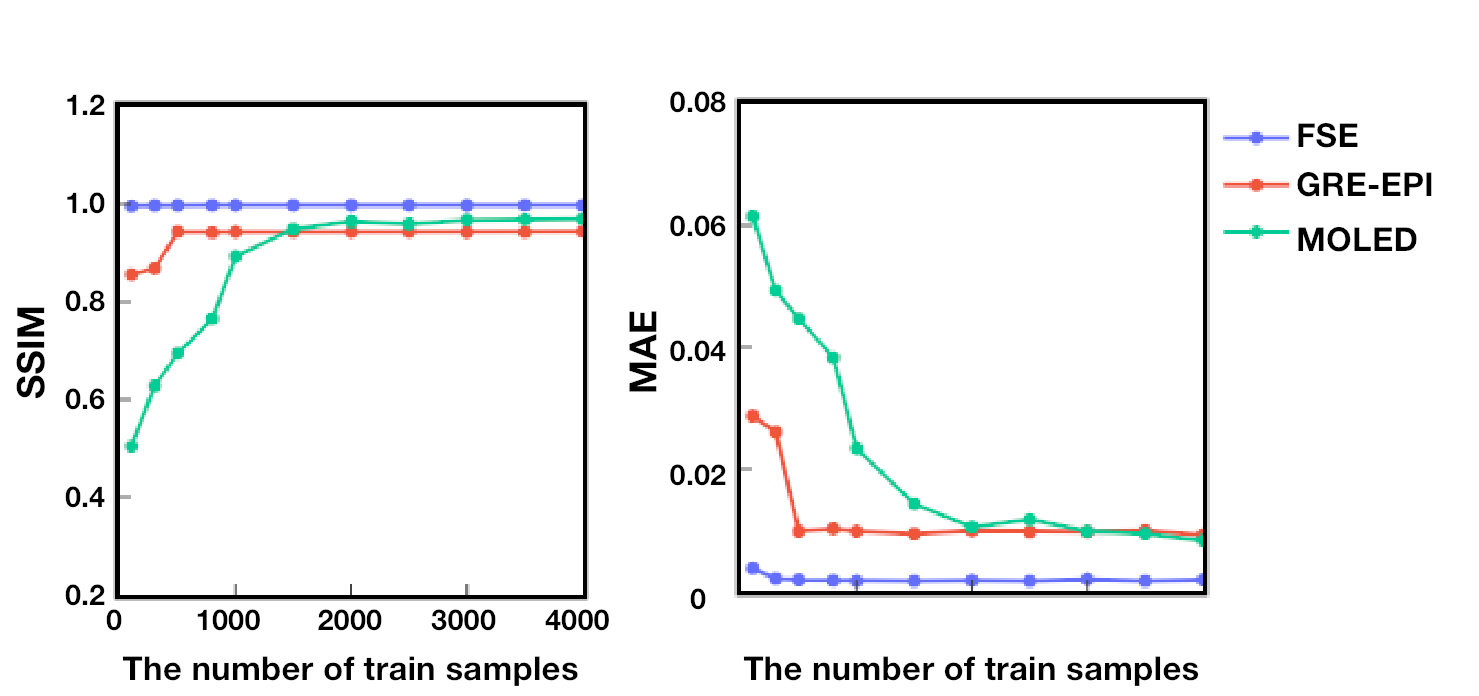}
	\captionsetup{labelfont=bf}
	\caption{The effects of the number of training samples on FSE (blue), GRE-EPI (red) and MOLED (green) simulation for Simu-Net.}
	\label{fig8}
\end{figure}

Recently, 3D MRI simulation of novel pulse sequences has become active due to increasing clinical applications. However, 3D Bloch simulation involves parallel computation of hundreds of times the spins in 2D simulation and requires gradient encoding in three directions. These factors limit the current simulation in relatively low resolution and take an extremely long computational time. Instead, Simu-Net does not involve explicit gradient encoding and FFT, but performs end-to-end image transformation in a data-driven manner. Thus, it can be applied to 3D MRI simulations by replacing all 2D operations with corresponding 3D counterparts. An preliminary experiment of the 3D bSSFP (balanced steady-state free precession) sequence was performed and illustrates the feasibility of 3D simulation with Simu-Net (see Supporting Information Figure S4).

As a proof-of-concept, Simu-Net has demonstrated the possibility of accelerating Bloch simulation using deep learning. Its applications may not be limited to synthetic data generation but also include large-scale high-accuracy simulation, model-driven self-supervised learning \cite{38liu2021magnetic}, pulse sequence optimization \cite{39loktyushin2021mrzero}, or reinforcement learning \cite{40jin2019self}. However, its limitations cannot be ignored, for example, variable spatial resolution (i.e., FOV and matrix size) and generalized models (adaptable to different pulse sequences) still remain challenges. Due to the relatively fixed number of nodes in a neural network, achieving variable output sizes may benefit from natural languages processing methods, such as recurrent neural network (RNN) or long short-term memory (LSTM) \cite{41sabidussi2021recurrent, 42ottens2022deep}. Besides, the “out-of-distribution” issue is still an open problem of the current implementation (see Supporting Information Figure S5). Although considering a wider range of parameters when generating samples can enhance the generalization of the model, a stronger physics-constrained version is highly desired. Future work will explore broader applications of Simu-Net, and we believe these ideas can be extended to other areas to accelerate the computation of forward physical model.

\section{CONCLUSION}
\label{sec:conclusion}
A novel deep learning framework called Simu-Net was proposed to greatly speed up Bloch simulations and verified on three different sequences. The simulation and \textit{in vivo} results show the flexibility, reliability and efficiency of this framework. This work firstly demonstrates the possibility of deep learning for 2D forward physical process approximation and may play an important role in promoting the development and optimization of MRI pulse sequences, the analysis and research of imaging experimental results, and the DL-based MRI methods.

\section*{Acknowledgements}
\label{sec:acknowledgements}

This work was supported in part by National Natural Science Foundation of China under Grant 22161142024, Grant 82071913 and Grant U1805261,
in part by the National Key R\&D Program of China under Grant 2022YFC2402102, 
in part by the Science and Technology Project of Fujian Province of China under Grant 2021Y9154.

\section*{Ethical statement}
\label{sec:ethical}

One participant took part under the prior approval of the ethics committee of Xiamen University. This research was conducted in accordance with the principles embodied in the Declaration of Helsinki and in accordance with local statutory requirements. Informed consent of the participant was obtained before scans. No approval ID number was specified.

%% If you have bibdatabase file and want bibtex to generate the
%% bibitems, please use
%%
\bibliographystyle{elsarticle-num} 
\bibliography{cas-refs}

\begin{thebibliography}{10}
\expandafter\ifx\csname url\endcsname\relax
  \def\url#1{\texttt{#1}}\fi
\expandafter\ifx\csname urlprefix\endcsname\relax\def\urlprefix{URL }\fi
\expandafter\ifx\csname href\endcsname\relax
  \def\href#1#2{#2} \def\path#1{#1}\fi

\bibitem{1drobnjak2006development}
I.~Drobnjak, D.~Gavaghan, E.~S{\"u}li, J.~Pitt-Francis, M.~Jenkinson,
  Development of a functional magnetic resonance imaging simulator for modeling
  realistic rigid-body motion artifacts, Magnetic Resonance in Medicine 56~(2)
  (2006) 364--380.

\bibitem{2cai2008sprom}
C.~Cai, M.~Lin, Z.~Chen, X.~Chen, S.~Cai, J.~Zhong, Sprom--an efficient program
  for nmr/mri simulations of inter-and intra-molecular multiple quantum
  coherences, Comptes Rendus Physique 9~(1) (2008) 119--126.

\bibitem{3stocker2010high}
T.~St{\"o}cker, K.~Vahedipour, D.~Pflugfelder, N.~J. Shah, High-performance
  computing mri simulations, Magnetic Resonance in Medicine 64~(1) (2010)
  186--193.

\bibitem{4benoit2005simri}
H.~Benoit-Cattin, G.~Collewet, B.~Belaroussi, H.~Saint-Jalmes, C.~Odet, The
  simri project: a versatile and interactive mri simulator, Journal of Magnetic
  Resonance 173~(1) (2005) 97--115.

\bibitem{5yang2022physics}
Q.~Yang, Z.~Wang, K.~Guo, C.~Cai, X.~Qu, Physics-driven synthetic data learning
  for biomedical magnetic resonance: The imaging physics-based data synthesis
  paradigm for artificial intelligence, IEEE Signal Processing Magazine early
  access (2022).

\bibitem{6yang2022model}
Q.~Yang, Y.~Lin, J.~Wang, J.~Bao, X.~Wang, L.~Ma, Z.~Zhou, Q.~Yang, S.~Cai,
  H.~He, et~al., Model-based synthetic data-driven learning (most-dl):
  Application in single-shot $\rm t_2$ mapping with severe head motion using
  overlapping-echo acquisition, IEEE Transactions on Medical Imaging (2022).

\bibitem{7frangi2018simulation}
A.~F. Frangi, S.~A. Tsaftaris, J.~L. Prince, Simulation and synthesis in
  medical imaging, IEEE Transactions on Medical Imaging 37~(3) (2018) 673--679.

\bibitem{8cai2018single}
C.~Cai, C.~Wang, Y.~Zeng, S.~Cai, D.~Liang, Y.~Wu, Z.~Chen, X.~Ding, J.~Zhong,
  Single-shot $\rm t_2$ mapping using overlapping-echo detachment planar
  imaging and a deep convolutional neural network, Magnetic Resonance in
  Medicine 80~(5) (2018) 2202--2214.

\bibitem{9cohen2018mr}
O.~Cohen, B.~Zhu, M.~S. Rosen, Mr fingerprinting deep reconstruction network
  (drone), Magnetic Resonance in Medicine 80~(3) (2018) 885--894.

\bibitem{10gavazzi2020deep}
S.~Gavazzi, C.~A. van~den Berg, M.~H. Savenije, H.~P. Kok, P.~de~Boer, L.~J.
  Stalpers, J.~J. Lagendijk, H.~Crezee, A.~L. van Lier, Deep learning-based
  reconstruction of in vivo pelvis conductivity with a 3d patch-based
  convolutional neural network trained on simulated mr data, Magnetic Resonance
  in Medicine 84~(5) (2020) 2772--2787.

\bibitem{11chen2020vivo}
L.~Chen, M.~Sch{\"a}r, K.~W. Chan, J.~Huang, Z.~Wei, H.~Lu, Q.~Qin, R.~G.
  Weiss, P.~van Zijl, J.~Xu, In vivo imaging of phosphocreatine with artificial
  neural networks, Nature Communications 11~(1) (2020) 1--10.

\bibitem{12della2020deepspio}
G.~Della~Maggiora, C.~Castillo-Passi, W.~Qiu, S.~Liu, C.~Milovic, M.~Sekino,
  C.~Tejos, S.~Uribe, P.~Irarrazaval, Deepspio: Super paramagnetic iron oxide
  particle quantification using deep learning in magnetic resonance imaging,
  IEEE Transactions on Pattern Analysis and Machine Intelligence 44~(1) (2020)
  143--153.

\bibitem{13chen2022ultrafast}
X.~Chen, W.~Wang, J.~Huang, J.~Wu, L.~Chen, C.~Cai, S.~Cai, Z.~Chen, Ultrafast
  water-fat separation using deep learning-based single-shot mri, Magnetic
  Resonance in Medicine 87~(6) (2022) 2811--2825.

\bibitem{14loecher2021using}
M.~Loecher, L.~E. Perotti, D.~B. Ennis, Using synthetic data generation to
  train a cardiac motion tag tracking neural network, Medical Image Analysis 74
  (2021) 102223.

\bibitem{15liu2017fast}
F.~Liu, J.~V. Velikina, W.~F. Block, R.~Kijowski, A.~A. Samsonov, Fast
  realistic mri simulations based on generalized multi-pool exchange tissue
  model, IEEE Transactions on Medical Imaging 36~(2) (2017) 527--537.

\bibitem{16xanthis2013mrisimul}
C.~G. Xanthis, I.~E. Venetis, A.~Chalkias, A.~H. Aletras, Mrisimul: a gpu-based
  parallel approach to mri simulations, IEEE Transactions on Medical Imaging
  33~(3) (2013) 607--617.

\bibitem{17xanthis2019coremri}
C.~G. Xanthis, A.~H. Aletras, coremri: A high-performance, publicly available
  mr simulation platform on the cloud, PLoS One 14~(5) (2019) e0216594.

\bibitem{18rupp2012fast}
M.~Rupp, A.~Tkatchenko, K.-R. M{\"u}ller, O.~A. Von~Lilienfeld, Fast and
  accurate modeling of molecular atomization energies with machine learning,
  Physical Review Letters 108~(5) (2012) 058301.

\bibitem{19kasim2021building}
M.~Kasim, D.~Watson-Parris, L.~Deaconu, S.~Oliver, P.~Hatfield, D.~Froula,
  G.~Gregori, M.~Jarvis, S.~Khatiwala, J.~Korenaga, et~al., Building high
  accuracy emulators for scientific simulations with deep neural architecture
  search, Machine Learning: Science and Technology 3~(1) (2021) 015013.

\bibitem{20spurio2022cosmopower}
A.~Spurio~Mancini, D.~Piras, J.~Alsing, B.~Joachimi, M.~P. Hobson, Cosmopower:
  Emulating cosmological power spectra for accelerated bayesian inference from
  next-generation surveys, Monthly Notices of the Royal Astronomical Society
  511~(2) (2022) 1771--1788.

\bibitem{21raissi2019physics}
M.~Raissi, P.~Perdikaris, G.~E. Karniadakis, Physics-informed neural networks:
  A deep learning framework for solving forward and inverse problems involving
  nonlinear partial differential equations, Journal of Computational Physics
  378 (2019) 686--707.

\bibitem{22karniadakis2021physics}
G.~E. Karniadakis, I.~G. Kevrekidis, L.~Lu, P.~Perdikaris, S.~Wang, L.~Yang,
  Physics-informed machine learning, Nature Reviews Physics 3~(6) (2021)
  422--440.

\bibitem{23van2022physics}
R.~L. van Herten, A.~Chiribiri, M.~Breeuwer, M.~Veta, C.~M. Scannell,
  Physics-informed neural networks for myocardial perfusion mri quantification,
  Medical Image Analysis 78 (2022) 102399.

\bibitem{24sarabian2022physics}
M.~Sarabian, H.~Babaee, K.~Laksari, Physics-informed neural networks for brain
  hemodynamic predictions using medical imaging, IEEE Transactions on Medical
  Imaging (2022).

\bibitem{25balsiger2020learning}
F.~Balsiger, A.~Jungo, O.~Scheidegger, B.~Marty, M.~Reyes, Learning bloch
  simulations for mr fingerprinting by invertible neural networks, in:
  International Workshop on Machine Learning for Medical Image Reconstruction,
  Springer, 2020, pp. 60--69.

\bibitem{26hamilton2019machine}
J.~I. Hamilton, N.~Seiberlich, Machine learning for rapid magnetic resonance
  fingerprinting tissue property quantification, Proceedings of the IEEE
  108~(1) (2019) 69--85.

\bibitem{27yang2020game}
M.~Yang, Y.~Jiang, D.~Ma, B.~B. Mehta, M.~A. Griswold, Game of learning bloch
  equation simulations for mr fingerprinting, arXiv preprint arXiv:2004.02270
  (2020).

\bibitem{28Cai7837616}
C.~Cai, Y.~Zeng, Y.~Zhuang, S.~Cai, L.~Chen, X.~Ding, L.~Bao, J.~Zhong,
  Z.~Chen, Single-shot $\rm t_2$ mapping through overlapping-echo detachment
  (oled) planar imaging, IEEE Transactions on Biomedical Engineering 64~(10)
  (2017) 2450--2461.

\bibitem{29Liang}
D.~Liang, J.~Cheng, Z.~Ke, L.~Ying, Deep magnetic resonance image
  reconstruction: Inverse problems meet neural networks, IEEE Signal Processing
  Magazine 37~(1) (2020) 141--151.

\bibitem{30Agg8434321}
H.~K. Aggarwal, M.~P. Mani, M.~Jacob, Modl: Model-based deep learning
  architecture for inverse problems, IEEE Transactions on Medical Imaging
  38~(2) (2019) 394--405.

\bibitem{31zhangdodnet}
J.~Zhang, Y.~Xie, Y.~Xia, C.~Shen, Dodnet: Learning to segment multi-organ and
  tumors from multiple partially labeled datasets, in: 2021 IEEE/CVF Conference
  on Computer Vision and Pattern Recognition (CVPR), 2021, pp. 1195--1204.

\bibitem{32NIPS2016_8bf1211f}
X.~Jia, B.~De~Brabandere, T.~Tuytelaars, L.~V. Gool, Dynamic filter networks,
  in: D.~Lee, M.~Sugiyama, U.~Luxburg, I.~Guyon, R.~Garnett (Eds.), Advances in
  Neural Information Processing Systems, Vol.~29, Curran Associates, Inc.,
  2016.

\bibitem{33bloch1946nuclear}
F.~Bloch, Nuclear induction, Physical Review 70 (1946) 460--474.

\bibitem{34falk2019u}
T.~Falk, D.~Mai, R.~Bensch, {\"O}.~{\c{C}}i{\c{c}}ek, A.~Abdulkadir,
  Y.~Marrakchi, A.~B{\"o}hm, J.~Deubner, Z.~J{\"a}ckel, K.~Seiwald, et~al.,
  U-net: Deep learning for cell counting, detection, and morphometry, Nature
  Methods 16~(1) (2019) 67--70.

\bibitem{35schmidt2014new}
R.~Schmidt, L.~Frydman, New spatiotemporal approaches for fully refocused,
  multislice ultrafast 2d mri, Magnetic Resonance in Medicine 71~(2) (2014)
  711--722.

\bibitem{36zhang2019robust}
J.~Zhang, J.~Wu, S.~Chen, Z.~Zhang, S.~Cai, C.~Cai, Z.~Chen, Robust single-shot
  $\rm t_2$ mapping via multiple overlapping-echo acquisition and deep neural
  network, IEEE Transactions on Medical Imaging 38~(8) (2019) 1801--1811.

\bibitem{vaswani2017attention}
A.~Vaswani, N.~Shazeer, N.~Parmar, J.~Uszkoreit, L.~Jones, A.~N. Gomez,
  {\L}.~Kaiser, I.~Polosukhin, Attention is all you need, Advances in neural
  information processing systems 30 (2017).

\bibitem{37bernstein2004handbook}
M.~A. Bernstein, K.~F. King, X.~J. Zhou, Handbook of MRI pulse sequences,
  Elsevier, 2004.

\bibitem{43biswal1995functional}
B.~Biswal, F.~Zerrin~Yetkin, V.~M. Haughton, J.~S. Hyde, Functional
  connectivity in the motor cortex of resting human brain using echo-planar
  mri, Magnetic Resonance in Medicine 34~(4) (1995) 537--541.

\bibitem{44edelman1994qualitative}
R.~R. Edelman, B.~Siewert, D.~G. Darby, V.~Thangaraj, A.~C. Nobre, M.~M.
  Mesulam, S.~Warach, Qualitative mapping of cerebral blood flow and functional
  localization with echo-planar mr imaging and signal targeting with
  alternating radio frequency, Radiology 192~(2) (1994) 513--520.

\bibitem{45nyquist}
H.~Nyquist, Certain topics in telegraph transmission theory, Transactions of
  the American Institute of Electrical Engineers 47~(2) (1928) 617--644.

\bibitem{38liu2021magnetic}
F.~Liu, R.~Kijowski, G.~El~Fakhri, L.~Feng, Magnetic resonance parameter
  mapping using model-guided self-supervised deep learning, Magnetic Resonance
  in Medicine 85~(6) (2021) 3211--3226.

\bibitem{39loktyushin2021mrzero}
A.~Loktyushin, K.~Herz, N.~Dang, F.~Glang, A.~Deshmane, S.~Weinm{\"u}ller,
  A.~Doerfler, B.~Sch{\"o}lkopf, K.~Scheffler, M.~Zaiss, Mrzero-automated
  discovery of mri sequences using supervised learning, Magnetic Resonance in
  Medicine 86~(2) (2021) 709--724.

\bibitem{40jin2019self}
K.~H. Jin, M.~Unser, K.~M. Yi, Self-supervised deep active accelerated mri,
  arXiv preprint arXiv:1901.04547 (2019).

\bibitem{41sabidussi2021recurrent}
E.~R. Sabidussi, S.~Klein, M.~W. Caan, S.~Bazrafkan, A.~J. den Dekker,
  J.~Sijbers, W.~J. Niessen, D.~H. Poot, Recurrent inference machines as
  inverse problem solvers for mr relaxometry, Medical Image Analysis 74 (2021)
  102220.

\bibitem{42ottens2022deep}
T.~Ottens, S.~Barbieri, M.~R. Orton, R.~Klaassen, H.~W. van Laarhoven,
  H.~Crezee, A.~J. Nederveen, X.~Zhen, O.~J. Gurney-Champion, Deep learning
  dce-mri parameter estimation: Application in pancreatic cancer, Medical Image
  Analysis (2022) 102512.

\end{thebibliography}

%% else use the following coding to input the bibitems directly in the
%% TeX file.

% \begin{thebibliography}{00}

% %% \bibitem{label}
% %% Text of bibliographic item

% \bibitem{}

% \end{thebibliography}
%\includepdfmerge{Supplementary.pdf,1-4}
\end{document}